\documentclass[aps,superscriptaddress,amsmath,amssymb,twocolumn,amsfonts,floatfix, longbibliography]{revtex4-2}
\usepackage{mathtools}
\usepackage{braket}
\usepackage[dvipsnames]{xcolor}
\usepackage{float}
\usepackage{subfigure}
\usepackage{upgreek}
\usepackage{pgffor}
\usepackage{float}
\usepackage{silence}
\usepackage[colorlinks=true,linktoc=page,linkcolor=blue,citecolor=magenta,urlcolor=Fuchsia]{hyperref}
\usepackage[normalem]{ulem}
\usepackage{sidecap,tikz}
\usepackage{orcidlink}
\usepackage{physics}
\usepackage{multirow}

\newcommand{\beq}{\begin{equation}}
\newcommand{\eeq}{\end{equation}}
\newcommand{\bea}{\begin{eqnarray}}
\newcommand{\eea}{\end{eqnarray}}

\newcommand{\eqn}[1] {Eq.~(\ref{#1})}
\newcommand{\fig}[1]{Fig.~\ref{#1}}
\newcommand{\sect}[1]{Sec.~\ref{#1}}
\newcommand{\mylabel}[1]{\label{#1}}

\newcommand{\noin}{\noindent}

\newcommand{\non}{\nonumber}

\begin{document}
\title{Quantum geometry-driven photogalvanic responses in semi-Dirac systems}	

\author{Bristi Ghosh}
\email{a21ph09025@iitbbs.ac.in}
\affiliation{School of Basic Sciences, Indian Institute of Technology Bhubaneswar, Argul, Jatni, Khurda, Odisha 752050, India}
\author{{Malay Bandyopadhyay}}
\thanks{Jointly supervised this work}
\email{malay@iitbbs.ac.in}
\affiliation{School of Basic Sciences, Indian Institute of Technology Bhubaneswar, Argul, Jatni, Khurda, Odisha 752050, India}
\author{{Snehasish Nandy}}
\thanks{Jointly supervised this work}
\email{snehasish@phy.nits.ac.in}
\affiliation{Department of Physics, National Institute of Technology Silchar, Assam 788010, India}
	
\date{\today}

\begin{abstract}

The photogalvanic effect (PGE), a fundamental nonlinear optical phenomenon in non-centrosymmetric materials, generates direct photocurrent under polarized light. 
Using quantum kinetic theory within the relaxation-time approximation, we theoretically investigate the PGE as a probe of quantum geometry in anisotropic type-I and type-II semi-Dirac (SD) systems, characterized by distinct electronic structures. We systematically analyse various microscopic contributions to the PGE conductivity, including injection, shift, resonance, higher-order pole, and anomalous terms,  and emphasize their connections to different quantum geometric quantities, namely, Berry curvature, quantum metric, and metric connection. By studying the frequency and chemical-potential dependence of the PGE conductivity in SD systems, we find that the optical conductivities in the type-II case are significantly enhanced relative to those in type-I. For the circular PGE (CPGE), Berry-curvature-driven contributions remain qualitatively similar in both phases, whereas the linear PGE (LPGE) displays clear qualitative differences. In particular, the $xxx$ component of the shift conductivity in the type-II phase reverses sign upon tuning the perturbation parameter $\delta$, providing a direct signature of the Lifshitz transition. In contrast, other components remain sign-invariant, as in type-I SD systems. These combined CPGE and LPGE signatures provide an unambiguous distinction between the two SD phases. The predicted effects, realizable in TiO$_2$/VO$_2$ heterostructures, establish PGE as a sensitive probe of quantum geometry with potential applications in polarization-selective photodetection, optical rectification, and next-generation optoelectronic devices.

\end{abstract}
	
\maketitle

\section{Introduction}

\begin{figure}[t]
    \includegraphics[width=\columnwidth]{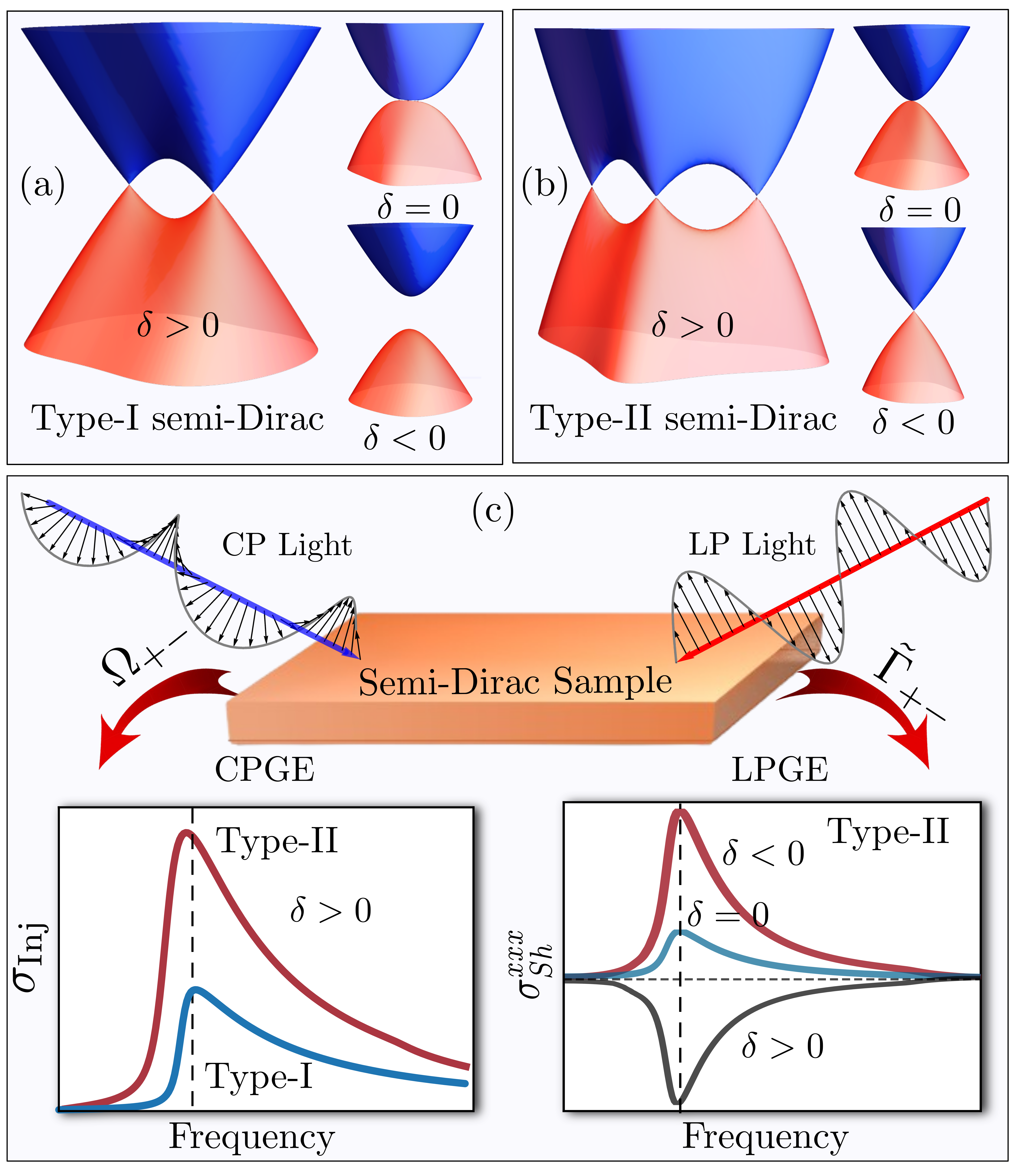}
    \caption{Schematic illustration of type-I and type-II semi-Dirac (SD) band structures and their corresponding photogalvanic responses. (a,b) Low-energy band dispersions for type-I and type-II SD systems for different values of the perturbation parameter $\delta$. For $\delta>0$, the type-I system hosts two Dirac nodes that merge at $\delta=0$ to form an SD node and become gapped for $\delta<0$. In contrast, the type-II system hosts three Dirac nodes for $\delta>0$, which merge at $\delta=0$ at the Brillouin-zone center, and the node moves away from the zone center without opening a gap for $\delta>0$. (c) Schematic of photogalvanic responses in a SD sample. Circularly polarized (CP) light probes the Berry curvature $\Omega_{+-}$, producing the circular photogalvanic effect (CPGE), whereas linearly polarized (LP) light probes the symplectic connection $\tilde{\Gamma}_{+-}$, which governs the shift current underlying the linear photogalvanic effect (LPGE). The bottom panels show representative frequency-dependent responses: the CPGE magnitude is markedly enhanced in the type-II phase relative to type-I, while the $xxx$ component of the shift conductivity changes sign across the three phases of type-II SD ($\delta>0,\delta=0,\delta<0,$), while no such sign change occurs in the type-I case. The vertical dashed line corresponds to twice the chemical potential.}
    \label{fig:schematic}
\end{figure}

The photogalvanic effect (PGE) refers to the generation of a dc photocurrent in a material that lacks inversion symmetry when it is illuminated~\cite{Belinicher_1980,Kraut_1979,Baltz_1981,Aversa_1995,Sipe_2000,fridkin_2001}. Depending on the polarization state of the incident light, the effect is classified as either the linear photogalvanic effect (LPGE) or the circular photogalvanic effect (CPGE). Recently, the PGE in quantum materials has attracted considerable attention, as it can be significantly more efficient than conventional photovoltaic mechanisms based on p–n junctions, making it promising for solar-cell applications. From a microscopic viewpoint, the PGE is governed by the quantum geometry of Bloch bands, which connects nonlinear optical responses to geometric quantities such as the Berry curvature, quantum metric, and their associated connections, including the metric and symplectic connections~\cite{Abigail_2024,Swati_2022,Jiang_2025,liu_2024,Hsu_2023}. Accordingly, most studies focus primarily on the injection~\cite{Aversa_1995,Sipe_2000,de_2017,Juan_2020,Ahn_2020,Watanabe_2021,Dai_2023,ahn_2022} and shift conductivity~\cite{Kraut_1979,Baltz_1981,Aversa_1995,Sipe_2000,Juan_2020,Ahn_2020,Watanabe_2021,Dai_2023,ahn_2022,Young_2012,Young_2012_Bi,Takahiro_2016,Kim_2017,Barik_2020,Yoshida_2025} mechanisms. In addition to the mechanisms of asymmetric carrier injection and real-space charge shifts during optical transitions, our work also considers Fermi-surface and Fermi-sea–related mechanisms~\cite{Gao_2021}, including anomalous, resonant, and higher-order pole contributions. These contributions are closely tied to underlying quantum-geometric quantities of the Bloch bands, such as the Berry curvature, quantum metric, metric connection, and symplectic connection.

The photogalvanic effect (PGE) has been extensively investigated in a variety of quantum materials, including Dirac and Weyl semimetals as well as magnetic semiconductors~\cite{Konig_2017,Steiner_2022,Taguchi_2016,Chan_2017,Sun_2017,ma2017,Ahn_2020,Watanabe_2021,Pankaj_2020,Gao_2020,Abigail_2024,Swati_2022}. Motivated by these developments, we turn our attention to semi-Dirac (SD) systems, which constitute a distinct class of two-dimensional materials characterized by highly anisotropic band dispersion-—linear in one momentum direction and parabolic in the orthogonal direction~\cite{Pardo_2009,Banerjee_2012,Shao_2024}. Such dispersions have been realized in a variety of platforms, including TiO$_2$/VO$_2$ multilayers, $\alpha$-(BEDT-TTF)$_2$I$_3$ salts under pressure~\cite{Katayama_2006}, phosphorene~\cite{Andres_2014,Rodin_2014}, and thin films of Cd$_3$As$_2$~\cite{Liu_2022}. Depending on the nature of their band topology, SD systems can be classified into two qualitatively distinct phases: type-I and type-II. The standard type-I SD model arises from the merging of two Dirac cones and hosts a topologically trivial band structure with zero Chern number, whereas the type-II SD model emerges from the merging of three Dirac cones and is characterized by a nonzero Chern number~\cite{Huang_2015}. In particular, recent works have examined the electronic and optical properties of two-dimensional SD systems, including their Floquet band structures~\cite{Islam2018,Chen2018}, optical conductivity~\cite{Carbotte2019,Sanderson_2018}, and nonlinear transport characteristics~\cite{Bristi2025}. Nevertheless, despite the extensive progress on linear optical and transport responses, the nonlinear optical properties—especially the photogalvanic effect (PGE) and its connection to quantum-geometric features—are still relatively underexplored.


In this work, we investigate the photogalvanic response of type-I and type-II SD systems, with particular emphasis on the role of quantum-geometric contributions beyond injection and shift conductivities. Using a quantum kinetic theory framework, we analyze the frequency dependence of the photogalvanic response to contrast the two semi-Dirac phases. We find that the responses of CPGE, governed by the Berry curvature, are strongly enhanced in the type-II SD phase compared to the type-I phase. In contrast, for LPGE, the symplectic connection plays a crucial role in determining the nature of the real part of the shift conductivity. For the type-I SD system, symmetry constraints allow only shift-conductivity components containing an even number of $x$ indices, yielding the nonvanishing components $\sigma_{\text{Sh}}^{yxx, xxy, xyx}$, while all components with an odd number of $x$ indices vanish. By contrast, the type-II SD system supports four nonzero shift-conductivity components, $\sigma_{\text{Sh}}^{xxx, xyy, yxy, yyx}$, each containing an odd number of $x$ indices, and the overall shift-conductivity response is significantly more pronounced than in the type-I phase. Notably, within the type-II SD system, the $\sigma_{\text{Sh}}^{xxx}$ component changes sign across the Lifshitz transition upon reversing the sign of the perturbative parameter $\delta$, whereas the remaining components do not exhibit a sign reversal under the sign change of $\delta$. These findings, which are experimentally accessible in TiO$_2$/VO$_2$ heterostructures, position the PGE as a sensitive probe of quantum geometry with promising applications in polarization-selective photodetection, optical rectification, and next-generation optoelectronic devices.

The remainder of the article is organized as follows. In \sect{sec:quantum_kinetic_theory_pge}, we present the quantum kinetic theory framework employed to study the PGE. The model Hamiltonians describing the type-I and type-II SD systems are introduced in \sect{sec:model}. In \sect{sec:results}, we discuss the main results of this work, with separate subsections devoted to the circular and linear photogalvanic responses: \sect{sec:cpge} analyzes the CPGE, while \sect{sec:lpge} presents the LPGE results, enabling a detailed comparison between the two SD phases. Finally, we summarize our findings and conclude in \sect{sec:conclusion}.

\section{Quantum Kinetic Theory of Photogalvanic Effect} \mylabel{sec:quantum_kinetic_theory_pge}


\begin{table*}[htb]
\centering
\label{tab:2nd_order_density_matrix}
\renewcommand{\arraystretch}{1.3}
\setlength{\tabcolsep}{6pt}
\begin{tabular}{|p{3.0cm}|p{9.5cm}|p{2.2cm}|p{0.5cm}|p{0.7cm}|}
\hline
\textbf{Category} & 
\textbf{Density matrix expression} & 
\textbf{Corresponding Conductivity} &
\multicolumn{2}{c|}{\textbf{Symmetry}} \\ \cline{4-5}
 & & & $\mathcal{T}$ & $\mathcal{PT}$ \\ \hline

Intraband–intraband  & 
$\rho_{dd}(\mathbf{k},t)=\frac{e^2}{\hbar^2} \, \tau g_{0}^{\omega} \, \partial_c \, \partial_b {\rho}_{mm}^{(0)} E_{0}^b E_{0}^c$ & 
$\sigma^{(2)}_{\text{Drude}}(\omega)$ & 
 0 & Finite\\ \hline

Intraband–interband  & 
$\rho_{do}(\mathbf{k},t)=\frac{e^2}{\hbar^2} \, \tau \, \sum_n \bigg( g_{nm}^{\omega} \mathcal{R}_{mn}^c \mathcal{R}_{nm}^b F_{nm} - g_{mn}^{\omega} \mathcal{R}_{mn}^b \mathcal{R}_{nm}^c F_{mn} \bigg) E_{0}^b E_{0}^c$ & 
$\sigma^{(2)}_{\text{Inj}}(\omega)$ & 
$\Omega$ & $\mathcal{G}$ \\ \hline

Interband–intraband  & 
$\rho_{od}(\mathbf{k},t)=i \frac{e^2}{\hbar^2} g_{mp}^0 g_{0}^{\omega} \mathcal{R}_{mp}^c \partial_b F_{mp} E_{0}^b E_{0}^c$ & 
$\sigma^{(2)}_{\text{Anm}}(\omega)$ & 
$\Omega$ & $\mathcal{G}$  \\ \hline

Interband–interband & 
$\begin{aligned}[t]
\rho_{oo}^{\text{I}}(\mathbf{k},t)&=i \frac{e^2}{\hbar^2} g_{mp}^0 \Big[
g_{mp}^{\omega} F_{mp} D_{mp}^c \mathcal{R}_{mp}^b \\
& - i \sum_{n \ne m \ne p} \big( g_{np}^{\omega} \mathcal{R}_{mn}^c \mathcal{R}_{np}^b F_{np} 
- g_{mn}^{\omega} \mathcal{R}_{mn}^b \mathcal{R}_{np}^c F_{mn} \big)\Big]E_{0}^b E_{0}^c
\end{aligned}$ & 
$\sigma^{(2)}_{\text{Sh}}(\omega)$ & 
$\tilde\Gamma$ & $\Gamma$ \\ \hline

Interband–interband  & 
$\rho_{oo}^{II}(\mathbf{k},t)=i \frac{e^2}{\hbar^2} g_{mp}^0 \mathcal{R}_{mp}^b g_{mp}^{\omega}\partial_c F_{mp}E_{0}^b E_{0}^c$ & 
$\sigma^{(2)}_{\text{Res}}(\omega)$ & 
$\Omega$ & $\mathcal{G}$  \\ \hline

Interband–interband  & 
$\rho_{oo}^{\text{III}}(\mathbf{k},t)=i \frac{e^2}{\hbar^2} g_{mp}^0 \mathcal{R}_{mp}^b \partial_c g_{mp}^{\omega} F_{mp}E_{0}^b E_{0}^c$ & 
$\sigma^{(2)}_{\text{HOP}}(\omega)$ & 
$\Omega$ & $\mathcal{G}$  \\ \hline
\end{tabular}
\caption{ Systematic classification of second-order density-matrix contributions to photogalvanic responses: corresponding conductivity expressions and symmetry requirements. The first column classifies second-order density-matrix terms according to their intraband or interband origin. The second column gives the quantum-mechanical expressions for these second-order terms. Double subscripts (dd, do, od, oo) denote the structure of the contributions: the first letter specifies whether the second-order density matrix is diagonal (d) or off-diagonal (o), and the second letter indicates whether it depends on the diagonal or off-diagonal component of the first-order density matrix. The conductivity column lists the corresponding conductivities, and the symmetry columns indicate which band geometric contributions remain finite under time-reversal ($\mathcal{T}$) or combined parity-time ($\mathcal{PT}$) symmetry constraints.}
\label{tab:2nd_order_density_matrix}
\end{table*}

The microscopic origin of the photogalvanic effect (PGE) can be systematically described within the framework of quantum kinetic theory, which captures the nonequilibrium carrier dynamics under light irradiation. In this approach, the underlying quantum geometry of Bloch states, encoded in the Berry curvature and the quantum metric, emerges naturally from the momentum-space evolution of the single-particle density matrix $\rho(\mathbf{k},t)$. The dynamics of $\rho(\mathbf{k},t)$ are governed by the quantum kinetic equation as~\cite{Sipe_2000}, 

\begin{equation}
\frac{d\rho(\mathbf{k},t)}{dt} + \frac{i}{\hbar}[\mathcal{H}(\mathbf{k},t),\rho(\mathbf{k},t)] = 0\,, \label{eq:dynamics}
\end{equation}
where $\mathcal{H}(\mathbf{k},t) = \mathcal{H}_0(\mathbf{k}) + \mathcal{H}_E+ \mathcal{H}_U$ is the full Hamiltonian, with $\mathcal{H}_0$ being the unperturbed band Hamiltonian. Here, $\mathcal{H}_U$ represents the disorder term of the Hamiltonian. The light–matter interaction is included in the length gauge via $\mathcal{H}_E=e\mathbf{r} \cdot \mathbf{E}(t)$, and the homogeneous external electric field is expressed as $\mathbf{E}(t) = \mathbf{E}_0 e^{-i\omega t}$+ c.c., where $\mathbf{E}_0=(E_{0}^x, E_{0}^y, E_{0}^z)$ specifies the field amplitude and $\omega$ is the driving frequency. In the
weak disorder limit, and within the relaxation time approximation, \eqn{eq:dynamics} can be rewritten as:
\beq
\frac{\partial \rho(\mathbf{k},t)}{\partial t} + \frac{i}{\hbar} \big[\mathcal{H}_0(\mathbf{k}) + \mathcal{H}_E, \rho(\mathbf{k},t) \big] + \frac{ \rho(\mathbf{k},t) - \rho^{(0)}(\mathbf{k},t) }{\tau}= 0\,. \label{eq:quantum_kinetic_equation} 
\eeq

\noin The third term describes relaxation toward the equilibrium distribution $\rho^{(0)}$, with $\tau$ denoting the relaxation time, which encapsulates the effects of various types of interactions and impurity effects and therefore depends on momentum, frequency, and temperature~\cite{Ghosh_2024}. We solve the kinetic equation perturbatively by expanding the single-particle density matrix in powers of the electric field as: $\rho = \rho^{(0)} + \rho^{(1)} + \rho^{(2)} \cdots$, with the $N$-th order term scaling as $\rho^{(N)} \sim |\mathbf{E}|^N$. Writing \eqn{eq:quantum_kinetic_equation} in the eigenbasis of $\mathcal{H}_0$, the recursive relation for $\rho^{(N)}$ reduces to

\begin{align}
\frac{\partial \rho^{(N)}_{mp}}{\partial t} + \frac{i}{\hbar} \big[\mathcal{H}_0, \rho^{(N)} \big]_{mp} + \frac{\rho^{(N)}_{mp}}{\tau} &=  \frac{e \mathbf{E}}{\hbar} \cdot \big[ \mathcal{D}_k\rho^{(N-1)} \big]_{mp}, \label{eq:Nth_order_kinetic}
\end{align}
\noin where the covariant derivative is defined as $\big[ \mathcal{D}_k\rho \big]_{mp} = \partial_k \rho_{mp} - i \big[\mathcal{R},\rho \big]_{mp}$. Here, the band-resolved Berry connection is defined as $\mathcal{R}_{mp} (\mathbf{k})= i \bra{u^{m}_{\mathbf{k}}}  \partial_{\mathbf{k}} \ket{u^{p}_{\mathbf{k}}}$, with $\ket{u^{m}_{\mathbf{k}}}$ denoting the Bloch wavefunction of the $m$-th band ~\cite{Sipe_1995,Pankaj_2022}. 
The first-order density matrix is obtained by setting $N=1$ in \eqn{eq:Nth_order_kinetic}:


\beq
\rho_{mp}^{(1)} = \frac{e}{\hbar} \sum_{\substack{\omega=\pm \omega_1}} g_{mp}^{\omega} \big(\partial_b \rho_{mp}^{(0)} \delta_{mp} + i \mathcal{R}_{mp}^b F_{mp} \big) E_{0}^b e^{- i \omega t}, \label{eq:first order density matrix}
\eeq

\noin where the first term arises from the diagonal density matrix and describes intraband dynamics, while the second term, involving the Berry connection ($\mathcal{R}$), represents the off-diagonal (interband) contribution capturing light-induced band coherence. The factor $g_{mp}^\omega = [1/\tau -i (\omega - \omega_{mp})]^{-1}$  describes the disorder-broadened joint density of states with $ \omega_{mp} = \frac{\epsilon_{m,\mathbf{k}}-\epsilon_{p, \mathbf{k}}}{\hbar}$. The summation over $\omega$ (i.e, $\omega_1$, $-\omega_1$) arises from the contribution of the electric field and its complex conjugate part. Using \eqn{eq:Nth_order_kinetic}, the second-order density matrix can be expressed in terms of $\rho_{mp}^{(1)}$ as 


\begin{align}
&\rho_{mp}^{(2)} \! = \non \\
\!\! &\frac{e}{\hbar} \!\!\!\!\sum_{\substack{\omega^\prime =\pm \omega_2 \\ \omega =\pm \omega_1}} \!\!\!\!\! g_{mp}^{(\omega + \omega^\prime)} \!\bigg[ \partial_c \rho_{mp}^{(1)} \!-i\! \sum_n \!\! \big( \mathcal{R}_{mn}^c \rho_{np}^{(1)} \!-\! \rho_{mn}^{(1)}\mathcal{R}_{np}^c \big)  \bigg] E_{0}^c e^{- i \omega^\prime t} \,.\label{eq:second order density matrix}
\end{align}

\noin The second-order density matrix gives rise to both the second-harmonic and photogalvanic (dc) currents. Specifically, when $\omega_1 = \omega_2$, it produces the second-harmonic current; the condition $\omega_1=-\omega_2$ corresponds to the photogalvanic current. The second-order contributions responsible for the photogalvanic response, encompassing both intraband and interband processes along with their respective conductivities, are summarized in Table~\ref{tab:2nd_order_density_matrix}. \\

The photogalvanic current can be expressed as,


\beq
j_{a}^{(2)} (0) = \sum_{\substack{\omega= \pm \omega}} \sigma^{abc}(\omega, -\omega) E^{b}(\omega) E^{c}(-\omega)\,, \label{eq:pge}
\eeq

\noin where $\sigma^{abc}(\omega, -\omega)$ denotes the symmetrized second-order photogalvanic conductivity tensor ($\sigma^{abc}(\omega, -\omega) = \sigma^{acb}(-\omega, \omega)$)~\cite{liu_2024}. Using symmetry arguments, the photogalvanic response in \eqn{eq:pge} can be decomposed into real and imaginary components:

\bea
 j_{a}^2 (0)&=&  2~\sigma^{abc}_{\text{Re}}(\omega, -\omega) E^{b}(\omega) E^{c}(-\omega) \non \\ 
 &+& 2i~\eta^{ad}(\omega, -\omega) (\mathbf{E}(\omega) \times \mathbf{E}(-\omega))_d,\,\label{eq:pge_response}
 \eea 
 
\noin where $(\mathbf{E}(\omega) \times \mathbf{E}(-\omega))_d = \epsilon_{dbc} E^{b} (\omega) E^{c} (-\omega)$ and $\sigma^{abc}_{\text{Im}}(\omega, -\omega)\equiv\eta^{ad}(\omega, -\omega) \epsilon_{dbc}$, with $\epsilon_{dbc}$ being the Levi-Civita tensor. The real part of the conductivity \eqn{eq:pge_response} gives the linear photogalvanic effect (LPGE), since the helicity $\mathbf{E} \times \mathbf{E}^*$ vanishes for linearly polarized light. In contrast, for circular polarization, the imaginary part of the conductivity remains finite, giving rise to the circular photogalvanic effect (CPGE). The CPGE current can be expressed as the difference between the responses to left- and right-circularly polarized light, $j^2_{CPGE} \equiv j^2_{LC} - j^2_{RC}$, which is proportional to the imaginary part of the conductivity~\cite{liu_2024,Jiang_2025}.

Now the second-order dc photocurrent can be obtained using \eqn{eq:second order density matrix} as

\bea
\mathbf{j^{(2)}(0)} = -e \mathrm{Tr}[\rho \hat{\mathbf{v}}],
\eea

\noin where the velocity operator in the eigenbasis of $\mathcal{H}_0$ reads $v_{pm}^a = \hbar^{-1} (\delta_{pm} \partial_a \epsilon_m + i \hbar~\omega_{pm} \mathcal{R}_{pm}^a)$. It includes the intraband term in the form of the band velocity and the interband term depending on the band-resolved Berry connection. Using the second-order density matrix expressions provided in Table 1, the total photogalvanic conductivity naturally decomposes into six distinct contributions: Drude, injection, shift, resonance, and anomalous terms,

\beq
\sigma^{abc} = \sigma^{abc}_{\mathrm{Drude}} + \sigma^{abc}_{\mathrm{Inj}} + \sigma^{abc}_{\mathrm{Res}} + \sigma^{abc}_{\mathrm{Anm}} + \sigma^{abc}_{\mathrm{HOP}} + \sigma^{abc}_{\mathrm{Sh}}\,,
\eeq
\noin where, 

\begin{widetext}
\bea
\sigma_{\text{Drude}}^{abc} &=& -\frac{e^3}{\hbar^2} \, \tau \int \, \frac{d^d k}{(2\pi)^d} g_{0}^\omega \sum_{m} \partial_a \omega_m \partial_c \partial_b f_{m}^{(0)} + (b, \omega) \leftrightarrow (c,-\omega) \label{eq:drude}\,, \\
\sigma_{\text{inj}}^{abc} &=& \frac{e^3}{\hbar^2} \, \tau \int \, \frac{d^d k}{(2\pi)^d} \sum_{p,m} \Delta^{a}_{mp} g_{mp}^{\omega} \bigg(\mathcal{G}_{mp}^{cb}-i \frac{\Omega_{mp}^{cb}}{2}\bigg) F_{mp} + (b, \omega) \leftrightarrow (c,-\omega) \label{eq:inj} \,, \\
\sigma^{abc}_{\text{Anm}}&=& - \frac{e^3}{\hbar^2} \int \, \frac{d^d k}{(2\pi)^d} g_{0}^{\omega}  \sum_{m,p} \omega_{mp} g_{mp}^0 \bigg(\mathcal{G}_{mp}^{ac}-i \frac{\Omega_{mp}^{ac}}{2}\bigg) \partial_b F_{mp}+ (b, \omega) \leftrightarrow (c,-\omega) \label{eq:anm} \,, \\
\sigma^{abc}_{\text{Res}}&=& - \frac{e^3}{\hbar^2} \int \, \frac{d^d k}{(2\pi)^d}   \sum_{m,p} \omega_{mp} g_{mp}^{\omega} g_{mp}^0 \bigg(\mathcal{G}_{mp}^{ab}-i \frac{\Omega_{mp}^{ab}}{2}\bigg) \partial_c F_{mp}+ (b, \omega) \leftrightarrow (c,-\omega) \label{eq:res} \,, \\
\sigma^{abc}_{\text{HOP}}&=& - \frac{e^3}{\hbar^2} \int \, \frac{d^d k}{(2\pi)^d}   \sum_{m,p} \omega_{mp}  g_{mp}^0 \bigg(\mathcal{G}_{mp}^{ab}-i \frac{\Omega_{mp}^{ab}}{2}\bigg) [\partial_c g_{mp}^{\omega}] F_{mp}+ (b, \omega) \leftrightarrow (c,-\omega) \label{eq:hop} \,, \\
\sigma^{abc}_{\text{Sh}}&=&  \frac{e^3}{\hbar^2} \int \, \frac{d^d k}{(2\pi)^d}   \sum_{m,p} \omega_{mp} g_{mp}^0 g_{mp}^\omega (\Gamma_{pm}^{bac}-i \Tilde{\Gamma}_{pm}^{bac}) F_{mp}+ (b, \omega) \leftrightarrow (c,-\omega) \label{eq:shift} \,.
\eea
\end{widetext}

\noin Here we define $\Delta^{a}_{mp}=v_{mm}^{0a}-v_{pp}^{0a}$, and $F_{mp}=f_{m}^{(0)} - f_{p}^{(0)}$, where $f_{m}^{(0)} \equiv \rho_{mm}^{(0)} \equiv [1+e^{\beta(\epsilon_{m,\mathbf{k}} - \mu)}]^{-1}$ is the Fermi-Dirac distribution function, with $\beta = 1 /(k_B T)$, $T$ is the absolute temperature and $\mu$ being the chemical potential.
The band geometric quantities $\mathcal{G}_{mp}^{ab}$, $\Omega_{mp}^{ab}$, $\Gamma_{mp}^{abc}$, and $\tilde{\Gamma}_{mp}^{abc}$ denote the band-resolved quantum metric and Berry curvature, metric connection and symplectic connection, respectively, defined as

\begin{align}
\mathcal{G}_{mp}^{ab} &= \mathrm{Re} [\mathcal{R}_{pm}^{a} \mathcal{R}_{mp}^{b}]\,, & \Omega_{mp}^{ab} &= -2 \mathrm{Im} [\mathcal{R}_{pm}^{a} \mathcal{R}_{mp}^{b}] \,, \non\\
\Gamma_{mp}^{abc} &= \text{Re}[\mathcal{R}_{pm}^{a} \mathcal{D}_{mp}^{b} \mathcal{R}_{mp}^{c}]\,, & \tilde{\Gamma}_{mp}^{abc} &= -\text{Im}[\mathcal{R}_{pm}^{a} \mathcal{D}_{mp}^{b} \mathcal{R}_{mp}^{c}] \,, \non
\end{align}
\vspace{-0.6cm}
\begin{equation}
\!\!\!\! \mathcal{R}_{pm}^{a} \mathcal{D}_{mp}^{b} \mathcal{R}_{mp}^{c}=-\frac{\bra{p}\partial_a \mathcal{H}\ket{m}}{\omega_{mp}^2} \bigg[ \frac{v_{mp}^b\Delta_{mp}^c+v_{mp}^c\Delta_{mp}^b}{\omega_{mp}} - w_{mp}^{bc} \bigg]\,.
\end{equation}

\noin where $v_{mp}^b=\bra{m} \partial_b \mathcal{H} \ket{p}$, and $w_{mp}^{bc}=\bra{m} \partial_b \partial_c \mathcal{H} \ket{p}$.


The conductivities in Eq.~(\ref{eq:drude}--\ref{eq:shift}) can be classified as Fermi sea or Fermi surface contributions to the photogalvanic effect, depending on whether the integrand contains the Fermi function itself or its derivative.

The symmetries of the bulk materials play a crucial role in enabling a non-zero PGE response. A finite second-order response requires broken inversion symmetry ($\hat{\mathcal{P}}$), since $j^{(2)}_a$ changes sign under $\hat{\mathcal{P}}$ while $E^b E^c$ remains invariant, forcing the response to vanish in centrosymmetric materials. Different quantum-geometric contributions persist depending on the presence of time-reversal ($\mathcal{T}$) or combined $\mathcal{PT}$ symmetry, as summarized in Table~\ref{tab:2nd_order_density_matrix}. In $\mathcal{T}$-symmetric systems, $\Omega_{mp}^{ab}$ contributes to the conductivity, whereas in $\mathcal{PT}$-symmetric systems, $\mathcal{G}_{mp}^{ab}$ provides the dominant contribution. To analyze how any conductivity tensor contributes to the LPGE and CPGE responses, we separate its real and imaginary components for both $\mathcal{T}$- and $\mathcal{PT}$-symmetric cases. As an illustration, we consider the anomalous conductivity and rewrite \eqn{eq:anm} in the following form.

\begin{widetext}
\beq
\sigma^{abc}_{\text{Anm}} \!\! = \!\! \frac{-e^3}{\hbar^2} \!\!\!\! \int \!\! \frac{d^d k}{(2\pi)^d} \!\! \sum_{m,p} \!\! \omega_{mp} \bigg\{\bigg(\text{Im}[g_{mp}^0g_{0}^{\omega}]-i~\text{Re}[g_{mp}^0g_{0}^{\omega}] \bigg)\frac{\Omega_{mp}^{ac}}{2}+\bigg(\text{Re}[g_{mp}^0g_{0}^{\omega}]+i~\text{Im}[g_{mp}^0g_{0}^{\omega}] \bigg)\mathcal{G}_{mp}^{ac} \bigg \}\partial_b F_{mp}+ (b, \omega) \leftrightarrow (c,-\omega)
\eeq
\end{widetext}

\noin In this expression, only the first term survives in a $\mathcal{T}$-symmetric system because the conductivity is solely determined by $\Omega_{mp}^{ac}$; its real part contributes to the LPGE, while its imaginary part gives the CPGE. Similarly, in a $\mathcal{PT}$-symmetric system, the conductivity originates from the last term, whose real and imaginary components likewise generate the LPGE and CPGE responses, respectively. A similar decomposition can be carried out for the term for CPGE and LPGE for the $\mathcal{T}$ and $\mathcal{PT}$ symmetric systems of other conductivities. This procedure yields circular injection and linear shift current for a $\mathcal{T}$ symmetric system and linear injection and circular shift current for a $\mathcal{PT}$ symmetric system~\cite{Jiang_2025,Orenstein2021}.

\begin{figure*}[t]
    \includegraphics[width=\textwidth]{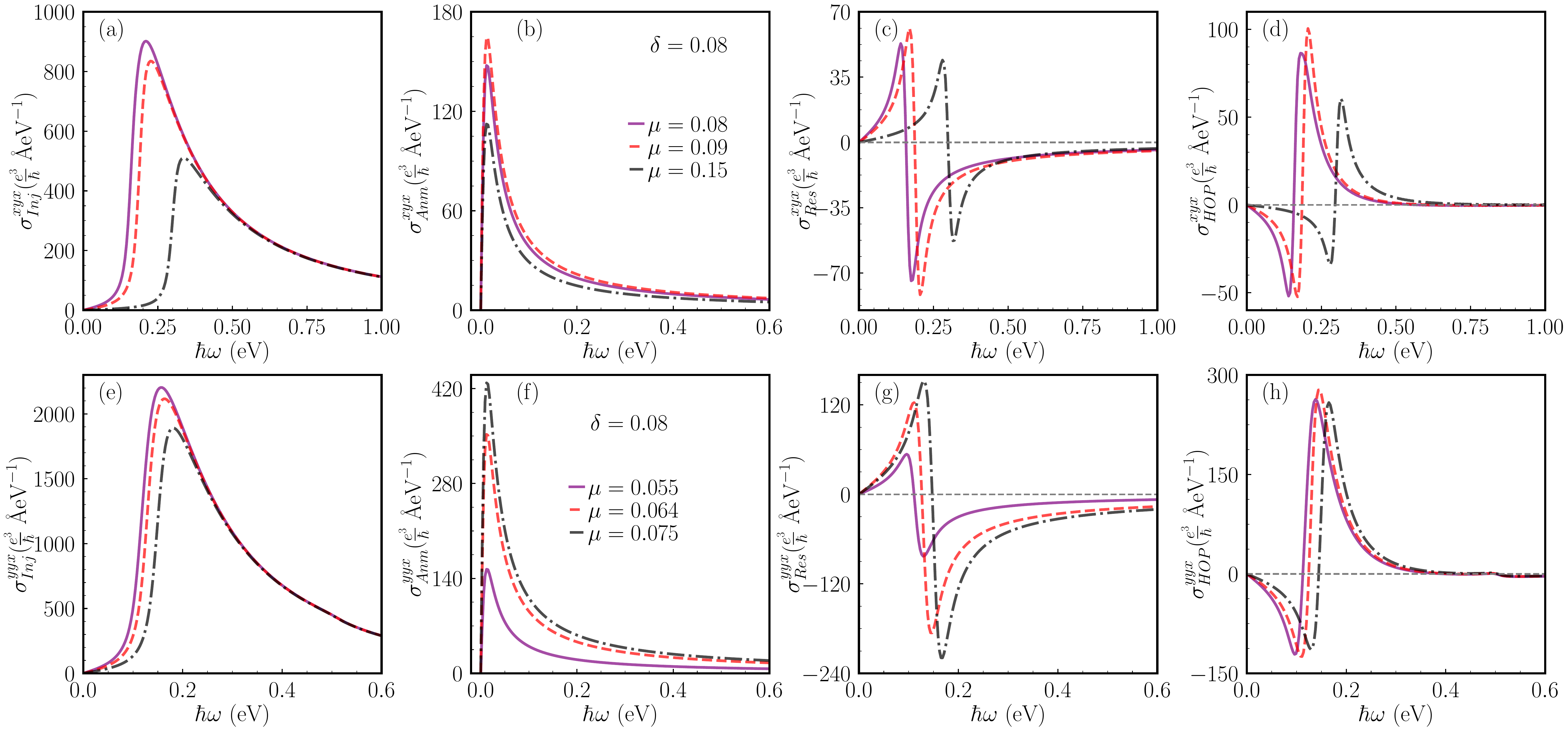}
    \caption{Circular photogalvanic conductivities are plotted as functions of photon frequency for several values of the chemical potential $\mu$. Panels (a–d) show the injection ($\sigma^{xyx}_{\text{Inj}}$), anomalous ($\sigma^{xyx}_{\text{Anm}}$), resonance ($\sigma^{xyx}_{\text{Res}}$), and higher-order-pole ($\sigma^{xyx}_{\text{HOP}}$) contributions for the type-I SD system, while panels (e–h) display the corresponding components for the type-II SD system. Injection and anomalous responses display a single prominent peak, while the resonance and higher-order-pole terms reveal two peaks. The model parameters used are ($\alpha$, $\hbar v_F$, $\Delta$, $T$, $\tau$) = ($0.75$ eV Å$^2$, $0.65$ eV Å, $0.05$ eV, 25 K, 0.05 ps) for both systems, with $\beta = 3.5$ eV Å$^2$ for the type-II SD system.}
    \label{fig:mu_var}
\end{figure*}



\section{PGE in type-I and type-II semi-dirac systems}

\subsection{Model} \mylabel{sec:model}


SD materials constitute a distinct class of two-dimensional electronic systems characterized by an anisotropic energy dispersion that exhibits linear behavior in one momentum direction and quadratic behavior in the orthogonal direction. The low-energy Hamiltonian describing such systems can be expressed in a generic form as
\begin{equation}
   H(\mathbf{k)} = \mathbf{d} (\mathbf{k}) \cdot \pmb{\sigma} \label{eq:ham}\,, 
\end{equation}

\noin where $\pmb{\sigma}$ are the Pauli matrices in pseudospin space and the vector $\mathbf{d}(\mathbf{k})$ encapsulates the momentum-dependent coefficients that determine the dispersion characteristics. Depending on its functional form, two distinct types of semi-Dirac dispersions emerge: type‑I and type‑II, each exhibiting a different kind of electronic behavior. The conventional type-I semi-Dirac system features two Dirac nodes, while the type-II semi-Dirac system hosts three Dirac nodes. The corresponding vector fields are $\mathbf{d}_1(\mathbf{k}) = (\alpha k_x^2 - \delta, \hbar v_F k_y, \Delta)$ for the type-I case and $\mathbf{d}_2(\mathbf{k}) = (\alpha k_x^2 - \hbar v_F k_y - \delta, \beta k_x k_y, \Delta)$ for the type-II case. Here, $\mathbf{k}$ is the crystal momentum with magnitude $k = \sqrt{k_x^2 + k_y^2}$, and the parameters are defined as follows: $\alpha = \frac{\hbar^2}{2m^*}$, where $m^*$ is the effective mass related to the $x$-direction,  $v_F$ is the Fermi velocity along the $y$-direction, and $\Delta$ represents the Dirac mass term. The parameter $\delta$ serves as a perturbation parameter that drives the transitions between different phases. For type-I, with $\Delta=0$ and $\delta > 0$, two gapless Dirac nodes appear at $(\pm\sqrt{\frac{\delta}{\alpha}}, 0)$. These nodes merge for $\delta=0$, where the dispersion assumes a semi-Dirac form. As $\delta$ becomes negative ($\delta < 0$), the system enters a trivial insulating phase. In contrast to the type‑I case, the type‑II semi-Dirac system with $\Delta=0$ hosts three distinct Dirac nodes in its low-energy spectrum for $\delta > 0$, located at $(\pm \sqrt{\delta/\alpha}, 0)$ and $(0, \delta/\hbar v_F)$. As $\delta$ decreases, these three nodes approach the Brillouin zone centre and merge into a single semi-Dirac point at $\textbf{k}=(0,0)$ when $\delta=0$. For $\delta<0$, the semi-Dirac node moves further away from the origin to (0,$\delta/\hbar v_F$) without opening the gap. The energy spectrum for type‑I and type‑II semi‑Dirac systems, incorporating a finite Dirac mass term, is expressed as,

\bea
\mathcal{E}^I_{\pm}(\mathbf{k}) &=& \pm \sqrt{(\alpha k_x^2 - \delta)^2 + \hbar^2 v^2_F k_y^2 + \Delta^2}\,,  \\
\mathcal{E}^{II}_{\pm}(\textbf{k}) &=& \pm \sqrt{(\alpha k_{x}^2 - \hbar v_F k_y - \delta)^2 + (\beta k_x k_y)^2 + \Delta^2} \,, \nonumber \\
\eea
where $ \pm $ denotes the conduction (+) and valence (-) bands, respectively. The energy dispersion for type-I and type-II SD is presented in \fig{fig:schematic}(a,b), with the plots corresponding specifically to the case of a vanishing mass term ($\Delta = 0$). Furthermore, both type-I and type-II systems exhibit Lifshitz transitions that can be induced by tuning $\delta$ at a fixed chemical potential $\mu$, or vice versa.




\begin{figure*}[t]
    \includegraphics[width=\textwidth]{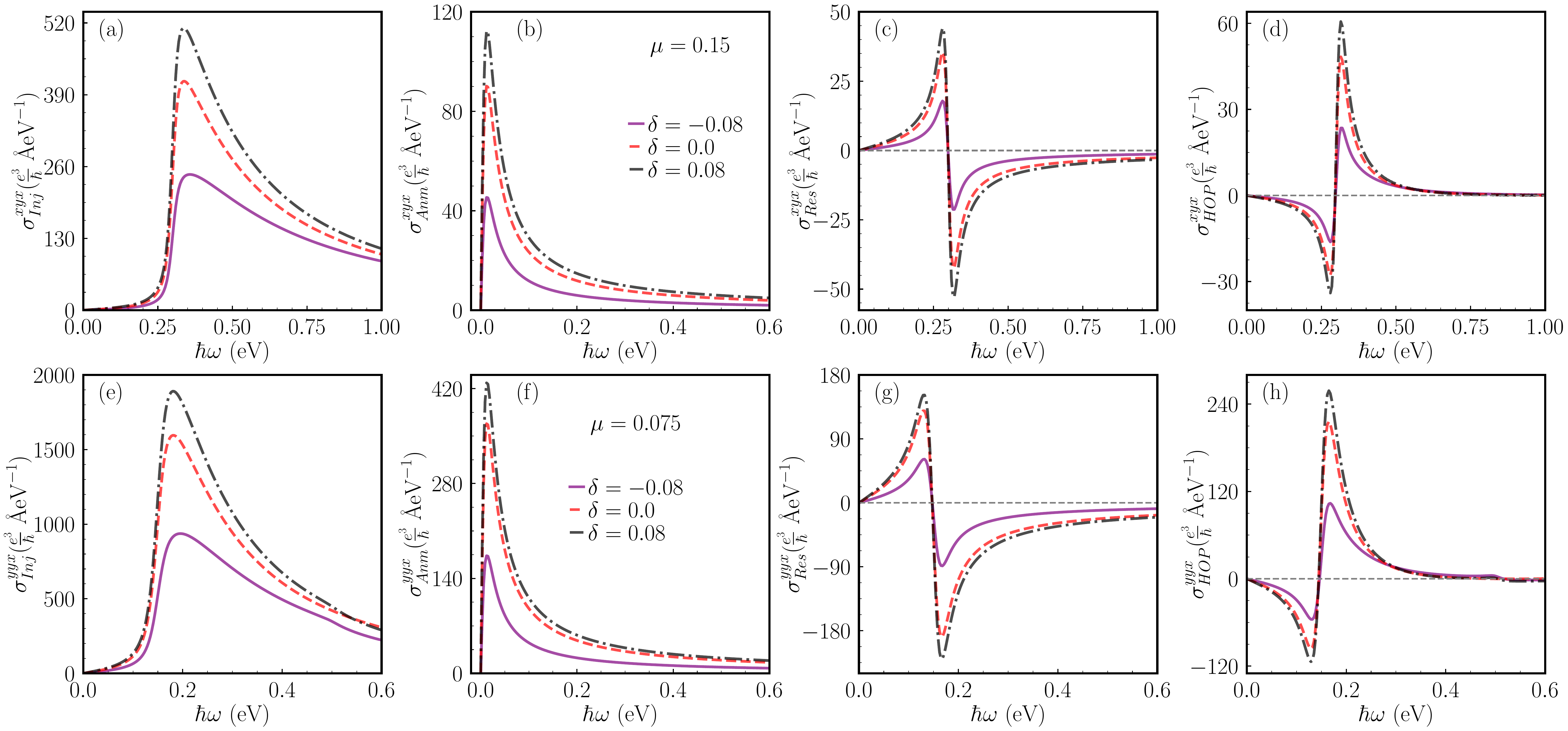}
    \caption{Dependence of circular photogalvanic conductivity on the photon frequency as the perturbation parameter $\delta$ is varied. The (a) injection ($\sigma^{xyx}_{\text{Inj}}$), (b) anomalous ($\sigma^{xyx}_{\text{Anm}}$), (c) resonance ($\sigma^{xyx}_{\text{Res}}$), and (d) higher-order-pole ($\sigma^{xyx}_{\text{HOP}}$) conductivities for the type-I semi-Dirac system are plotted. Panels (e)–(h) present the corresponding $\sigma^{yyx}$ components for the type-II semi-Dirac system. Notably, type-II displays systematically larger amplitudes for all contributions due to its additional Dirac node. Peak positions remain largely insensitive to $\delta$, while amplitudes adjust smoothly. The model parameters are the same as in \fig{fig:mu_var}.}
    \label{fig:delta_var}
\end{figure*}

A finite PGE arises only when inversion symmetry is broken. In the absence of the mass term $\Delta$, the Hamiltonian in \eqn{eq:ham} preserves inversion symmetry and forbids any dc photocurrent, whereas a finite $\Delta$ lifts this symmetry and enables nonlinear optical responses. Although both type-I and type-II SD systems preserve time-reversal symmetry, broken inversion symmetry generates nontrivial band geometry that governs the PGE response. In such time-reversal-symmetric systems, the PGE is fully controlled by the Berry curvature and the symplectic connection: the shift current originates from the symplectic connection, while the injection, anomalous, resonant, and higher-order-pole currents are governed by the Berry curvature.

For the type-I SD system, characterized by two merging Dirac nodes, the interband Berry curvature between the conduction ($+$) and valence ($-$) bands is given by
\beq
    \Omega_{+-}^{xy} = \frac{\alpha \hbar v_F \Delta k_x}{({\mathcal{E}^{I}(\mathbf{k})})^3}
\eeq

\noin For the type-II semi-Dirac system, hosting three Dirac nodes, the Berry curvature exhibits a distinct momentum dependence,
\beq
    \Omega_{+-}^{xy} =
    \frac{\Delta \left( 2 k_x^2 \alpha \beta + k_y v_F \beta \hbar \right)}
    {2 ({\mathcal{E}^{II}(\mathbf{k})})^3}
\eeq

\noin The closed-form expression for the symplectic connection in these semi-Dirac models is algebraically cumbersome and is therefore evaluated numerically in our analysis.

\subsection{Results and Discussions} \mylabel{sec:results}

In this section, we compare the nonlinear photogalvanic response of type-I and type-II semi-Dirac systems, highlighting how their distinct nodal configurations shape the optical conductivity. Both models share anisotropic linear–quadratic dispersion, but type-I hosts two merging Dirac nodes, whereas type-II contains three. This difference reorganizes the quantum geometry of the bands, including the Berry curvature, quantum metric, and related geometric connections, as well as the joint density of states and interband transition pathways, resulting in characteristic contrasts between their circular and linear photogalvanic responses. We analyze the frequency dependence and the role of chemical potential $(\mu)$ and perturbation parameter $(\delta)$ across the injection, anomalous, resonant, higher-order-pole, and shift-current contributions to identify the optical signatures tied to the underlying band geometry. 


\subsubsection{CPGE} \mylabel{sec:cpge}

We begin by analyzing the CPGE, which serves as a sensitive probe of interband coherence and Berry-curvature–driven optical processes under circularly polarized excitation. The imaginary part of the conductivity tensor contributes to CPGE responses. In the type-I SD system, the combined presence of mirror symmetry $\mathcal{M}_x$ and time-reversal symmetry $\mathcal{T}$ implies $\Omega^{xy}_{mp}(k_x,k_y)=\Omega^{xy}_{mp}(k_x,-k_y)$. As a result, the Brillouin-zone integrands for the CPGE components with the structure $\sigma^{\text{Inj/Anm/Res/HOP}}_{yyx}$ are odd under $k_y\to-k_y$ and therefore vanish. Consequently, the only independent nonvanishing CPGE tensor component is $\sigma^{\text{Inj/Anm/Res/HOP}}_{xyx}$. A similar argument applies to the type-II SD system, but here it arises from the combined action of the mirror symmetry $\mathcal{M}_y$ and $\mathcal{T}$ symmetry, which imposes $\Omega^{xy}_{mp}(k_x,k_y)=\Omega^{xy}_{mp}(-k_x,k_y)$. This symmetry again eliminates the $\sigma^{\text{Inj/Anm/Res/HOP}}_{xyx}$ components, leaving $\sigma^{\text{Inj/Anm/Res/HOP}}_{yyx}$ as the independent nonvanishing CPGE tensor element in the type-II case as well.

The individual CPGE contributions, corresponding to distinct microscopic channels of photocurrent generation, are shown in \fig{fig:mu_var}. It is clear from the \fig{fig:mu_var}(a, e) that the imaginary part of the injection conductivity $(\rm{Im}[\sigma_{\rm \text{Inj}}])$ depicts a pronounced frequency dependence in both type-I and type-II SD systems. The $\sigma_{\text{Inj}}^{xyx}$ component for the type-I SD system and the $\sigma_{\text{Inj}}^{yyx}$ component for the type-II SD system exhibit a sharp peak near $\hbar \omega \approx 2\mu$, with the peak position shifting systematically as the chemical potential $\mu$ is varied. Similar behavior of the injection conductivity has been reported in other systems~\cite{Ahn_2020,Pankaj_2022}. The primary distinction between the two models lies in the magnitude of the response: the type-II system produces an injection current nearly twice that of type-I, reflecting the enhanced phase space associated with its three-node topology.

\begin{figure*}[htb]
    \includegraphics[width=\textwidth]{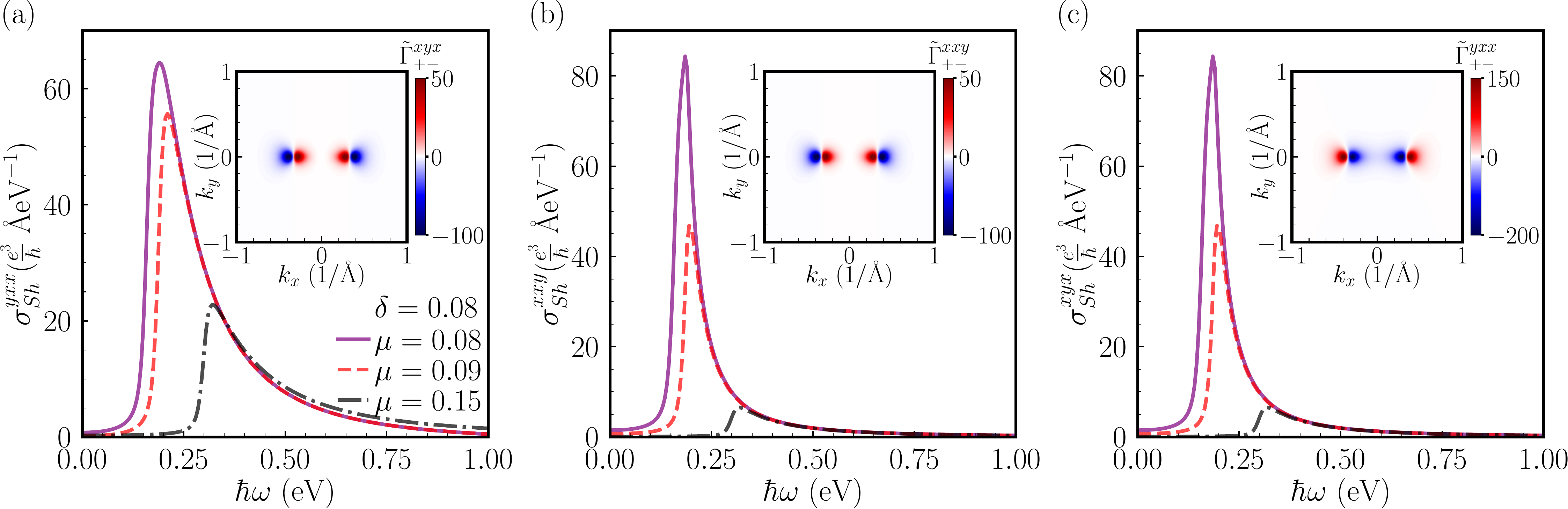}
    \caption{ Chemical potential ($\mu$) dependence of different components of LPGE shift conductivity for type-I semi-Dirac system as functions of photon frequency. The photon frequency dependence of (a) $\sigma^{yyx}_{Sh}$, (b) $\sigma^{xxy}_{Sh}$, and $\sigma^{xyx}_{Sh}$-components of shift conductivity is shown for multiple values of $\mu$. The components exhibit identical qualitative behaviour: a single negative peak whose position shifts to higher frequencies with increasing $\mu$. The model parameters used are ($\alpha$, $\hbar v_F$, $\Delta$, $T$, $\tau$) = ($0.75$ eV Å$^2$, $0.65$ eV Å, $0.05$ eV, 25 K, 0.1 ps).}
    \label{fig:type1_mu_var_shift}
\end{figure*}

\begin{figure*}[htb]
    \includegraphics[width=\textwidth]{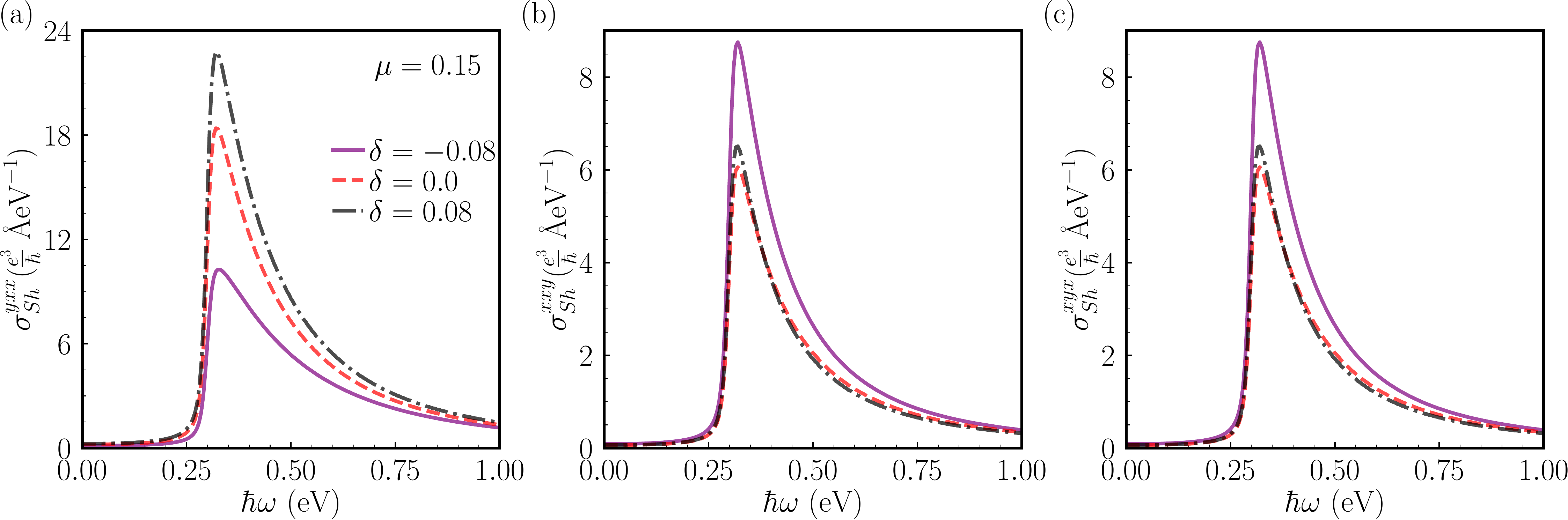}
    \caption{Perturbation parameter ($\delta$) dependence of LPGE shift current components for the type-I system is plotted for (a) $\sigma^{yyx}_{Sh}$, (b) $\sigma^{xxy}_{Sh}$, and (c) $\sigma^{xyx}_{Sh}$. Different components display identical spectral behaviour, with a single negative peak whose amplitude varies smoothly with $\delta$, while the peak position remains essentially unchanged. This reflects the fact that $\delta$ alters the nodal geometry and phase-space weighting without modifying the characteristic interband transition energy. The model parameters are the same as in \fig{fig:type1_mu_var_shift}.}
    \label{fig:type1_delta_var_shift}
\end{figure*}

The imaginary part of the anomalous conductivity $(\rm{Im}[\sigma_{\text{Anm}}])$ for both type-I and type-II SD systems is shown in \fig{fig:mu_var}(b,f). This contribution arises from Berry–curvature–weighted interband processes, with the dominant signal originating from regions near the SD nodes, where the Berry curvature is most strongly concentrated. As a result, the anomalous response is set by the intrinsic interband energy scale of the nodal dispersion, which fixes the peak position. Consequently, changes in the chemical potential affect only the spectral weight through $\partial_k F_{mp}$, while the position of the peak in frequency remains unchanged. Moreover, the factor $\mathrm{Re}[g^{\omega}_0 g^{0}_{mp}]$ reaches its maximum near $\hbar\omega \approx 0$, leading to a pronounced low-frequency peak in the anomalous conductivity, consistent with \fig{fig:mu_var}(b,f).

The imaginary component of the resonance conductivity, $\mathrm{Im}[\sigma_{\text{Res}}]$, reveals a rich frequency dependence, with a distinctive sign-changing (positive–negative) sequence, as shown in \fig{fig:mu_var}(e,g). This sign reversal follows directly from the structure of the integrand. The momentum derivative of the Fermi function leads to $-\delta(\mathcal{E}_m -\mu) \partial_{\mathbf{k}} \mathcal{E}_m + \delta(\mathcal{E}_p -\mu) \partial_{\mathbf{k}} \mathcal{E}_p$. For the case $\omega_m = -\omega_p$, when combined with the momentum derivative, the resonant factor $\mathrm{Re}[g^{\omega}_{mp} g^{0}_{mp}]$ crosses zero near $\hbar\omega \approx 2\mu$, the resulting frequency dependence exhibits the observed positive-to-negative sequence, consistent with earlier findings~\cite{Pankaj_2022}.

The higher-order-pole (HOP) contribution originates from the momentum derivative of the joint density of states, $\partial_{\mathbf{k}}g^{\omega}_{mp}$, weighted by Fermi-sea occupations. The imaginary part of the HOP conductivity, $\mathrm{Im}[\sigma_{\text{HOP}}]$, exhibits a characteristic dispersive lineshape with a negative-to-positive peak sequence, as shown in \fig{fig:mu_var}(d,h). The nature of the curve is governed by the interband spacing $\omega_{mp}$, since the factor $\mathrm{Re}[g^{0}_{mp} \partial_{\mathbf{k}}g^{\omega}_{mp}]$ crosses zero at $\omega \approx \omega_{mp}$. As the chemical potential is varied, the set of occupied states contributes through $F_{mp}$ shifts in momentum space. Because $\omega_{mp}(k)$ varies across the Brillouin zone, changes in $\mu$ alter the dominant contributing momenta and thereby shift the frequency at which the peak occurs. In addition, the peak magnitude of the HOP conductivity in the type-II SD system is approximately $1.5$ times larger than that of the type-I system, reflecting the enhanced interband phase space available in the three-node topology.

The most salient distinction between the CPGE responses of the two SD phases lies in their overall magnitude. The peak magnitudes of all conductivity components are substantially larger in the type-II case—typically about a factor of two—owing to the enhanced interband phase space associated with the additional Dirac node. This enhancement is further amplified by the underlying Berry-curvature geometry: while the type-I phase hosts two symmetry-related nodes whose Berry-curvature contributions partially cancel, the type-II phase contains three nodes together with a $k_x k_y$ mixing term that redistributes curvature more asymmetrically in momentum space. This reduces cancellations between opposite-momentum regions and increases the phase-space volume contributing with the same sign, thereby yielding a much larger CPGE response. 


\begin{figure*}[t]
    \includegraphics[width=\textwidth]{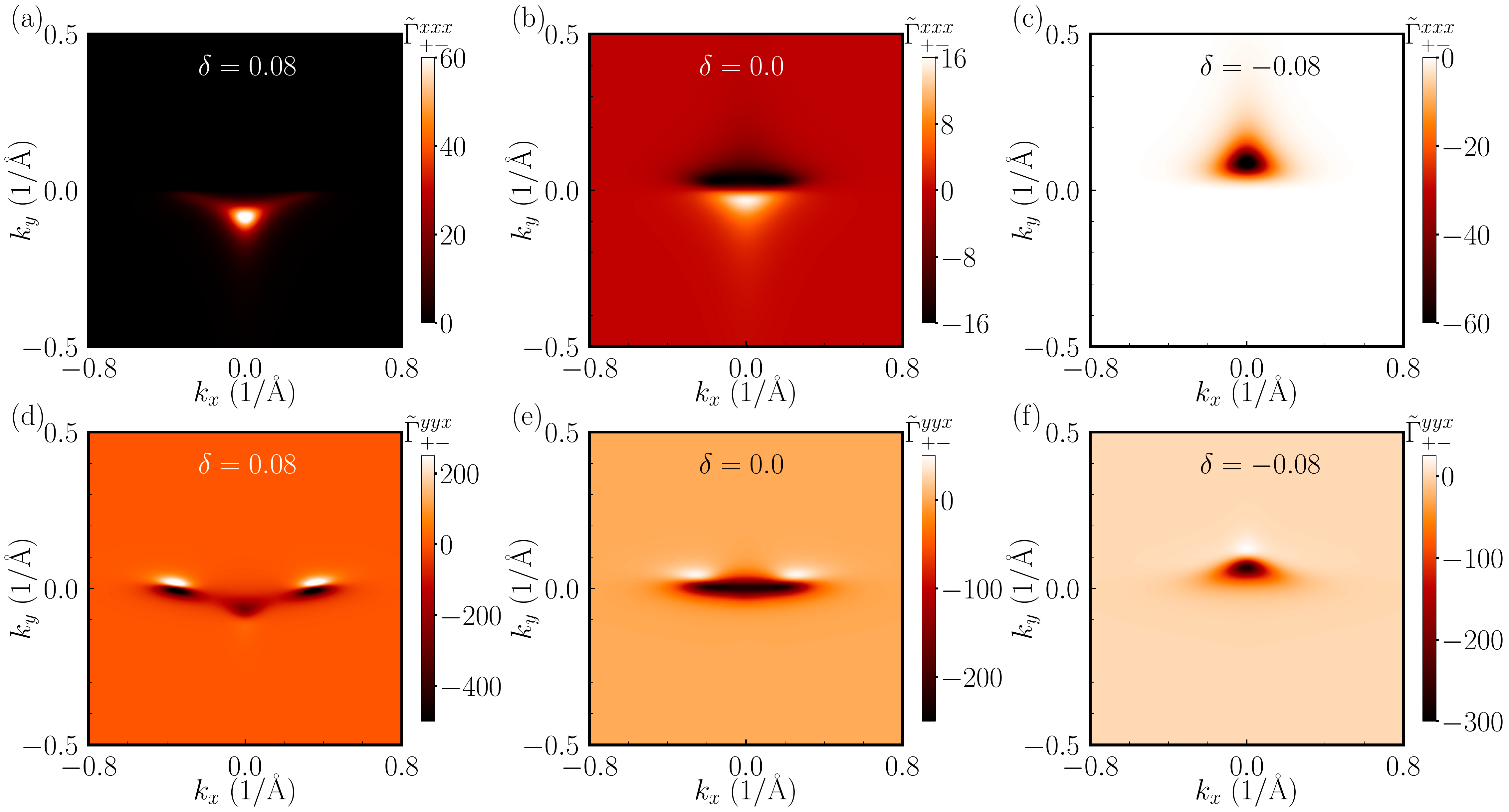}
    \caption{Momentum-space distributions of the symplectic connection in a type-II SD system. (a–c) $\tilde{\Gamma}^{xxx}_{+-}$ associated with the $xxx$ component of the real shift conductivity $\sigma^{xxx}_{\mathrm{Sh}}$ for $\delta=0.08$, $0.0$, and $-0.08$, respectively. (d–f) $\tilde{\Gamma}^{yyx}_{+-}$ corresponding to the $yyx$ component of the real shift conductivity $\sigma^{yyx}_{\mathrm{Sh}}$ for the same values of $\delta$. A clear sign reversal of $\tilde{\Gamma}^{xxx}_{+-}$ with $\delta$ is observed, which directly manifests in the frequency dependence of $\sigma^{xxx}_{\mathrm{Sh}}$.}
    \label{fig:type2_symp_connection}
\end{figure*}

\fig{fig:delta_var} shows that the frequency dependence of all CPGE contributions follows qualitatively similar trends across different perturbation parameters $(\delta)$ in both type-I and type-II semi-Dirac systems. As $\delta$ is varied, each system undergoes changes in its nodal topology, leading to corresponding shifts in the magnitude of the conductivities through the redistribution of Berry curvature and the modification of resonant $k$-space channels. The main difference remains quantitative: type-II consistently exhibits larger magnitudes for all contributions due to the presence of an additional Dirac node and the resulting enhancement of allowed interband transitions. Thus, although $\delta$ drives analogous spectral evolution in both systems, the systematically amplified response in type-II serves as a clear indicator of its richer anisotropic band structure.


\subsubsection{LPGE} \mylabel{sec:lpge}

Having discussed the circular photogalvanic response, we now turn to the LPGE, focusing in particular on the shift current, which originates from real-space carrier displacement associated with interband photoexcitation. Unlike the CPGE, where the imaginary part of the conductivity governs the response, the LPGE is determined by the real part of the nonlinear conductivity tensor. A key distinction between the two SD systems based on LPGE lies in the symmetry-allowed tensor structure: the type-I phase supports three independent nonvanishing shift-current components, whereas the type-II phase admits four, reflecting their differing symmetry constraints and nodal geometries.

\fig{fig:type1_mu_var_shift} shows all nonvanishing components of the real part of the shift conductivity in the type-I semi-Dirac system. Owing to the underlying mirror $\mathcal{M}_x$ and time-reversal $(\mathcal{T})$ symmetries, only tensor elements containing an even number of $x$ indices survive, yielding the allowed components $(yxx,xxy,xyx)$, while all components with an odd number of $x$ indices vanish. Each of the surviving components exhibits a single pronounced peak that systematically shifts toward higher frequencies as the chemical potential $\mu$ increases, accompanied by a monotonic reduction in its amplitude. This trend reflects the Fermi-surface–weighted nature of the shift current: increasing $\mu$ moves the active interband transition window to momenta with larger interband energy differences $\omega_{mp}$, thereby pushing the peak to higher frequencies. The nearly identical qualitative and quantitative responses of the $xxy$ and $xyx$ components arise from their sharing the same underlying symmetry structure of symplectic connection $(\tilde{\Gamma}^{abc}_{mp})$ in the type-I semi-Dirac Hamiltonian, whereas the $yxx$ component exhibits a distinct behaviour, differing in both the peak magnitude and its variation with $\delta$.

\fig{fig:type1_delta_var_shift} shows the frequency dependence of the real part of the shift conductivity for different values of $\delta$ in the type-I system. Varying $\delta$ alters only the amplitude of the nonvanishing shift-current components while leaving the peak position essentially unchanged. This behavior arises because $\delta$ tunes the separation between the two nodes but does not significantly modify the dominant interband energy scale of the type-I SD system. As a result, the shift-current peak occurs at nearly the same frequency for all $\delta$. Notably, the magnitude of the $yxx$ component decreases as $\delta$ is tuned from positive to negative values—corresponding to a transition from a two-node phase $(\delta >0 )$ through a single-node critical point $\delta = 0$ into a gapped phase $(\delta < 0)$—whereas the other two components, $xxy$ and $xyx$, display the opposite trend. This contrasting behavior is governed by the structure of the $\tilde{\Gamma}^{abc}_{mp}$, which redistributes the geometric contribution to the shift current among the symmetry-allowed tensor components as the nodal configuration evolves with $\delta$.

\begin{figure}[t]
    \includegraphics[width=\columnwidth]{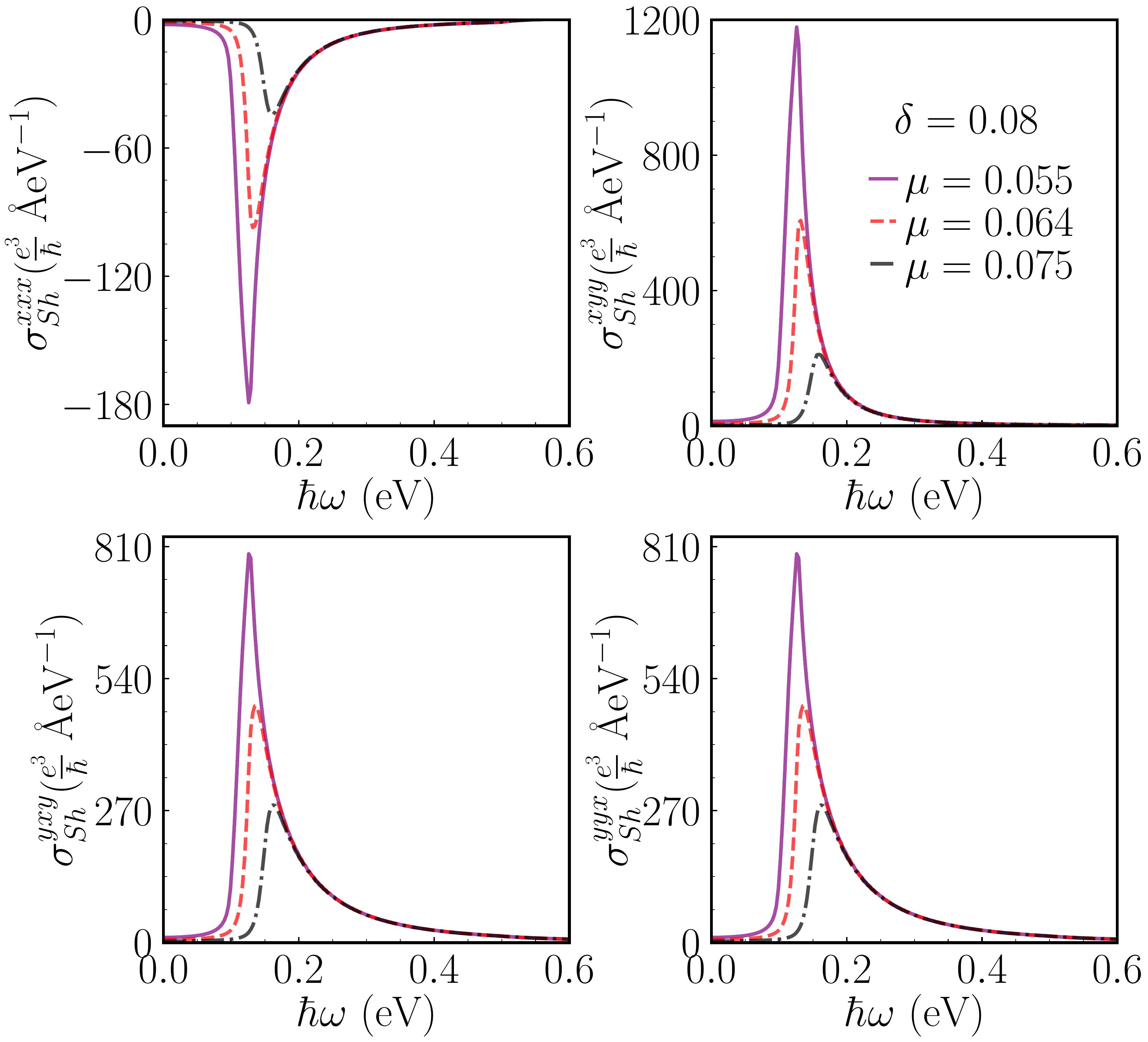}
    \caption{The shift current components (a) $\sigma^{xxx}_{Sh}$, (b)$\sigma^{xyy}_{Sh}$, (c)$\sigma^{yxy}_{Sh}$, and (d) $\sigma^{yyx}_{Sh}$ for type-II systems are shown as functions of photon frequency for several values of $\mu$. The $xxx$ component remains strictly positive, while the other three components stay negative across all $\mu$. Although the tensor elements differ in sign, they share similar peak positions that shift to higher frequencies with increasing $\mu$, reflecting the Fermi-level–driven movement of the active resonant transitions in the type-II band structure. The model parameters used are ($\alpha$, $\hbar v_F$, $\Delta$, $\beta$, $T$, $\tau$) = ($0.75$ eV Å$^2$, $0.65$ eV Å, $0.05$ eV, $3.5$ eV Å$^2$ , 25 K, 0.1 ps).}
    \label{fig:type2_mu_var}
\end{figure}

\begin{figure}[t]
    \includegraphics[width=\columnwidth]{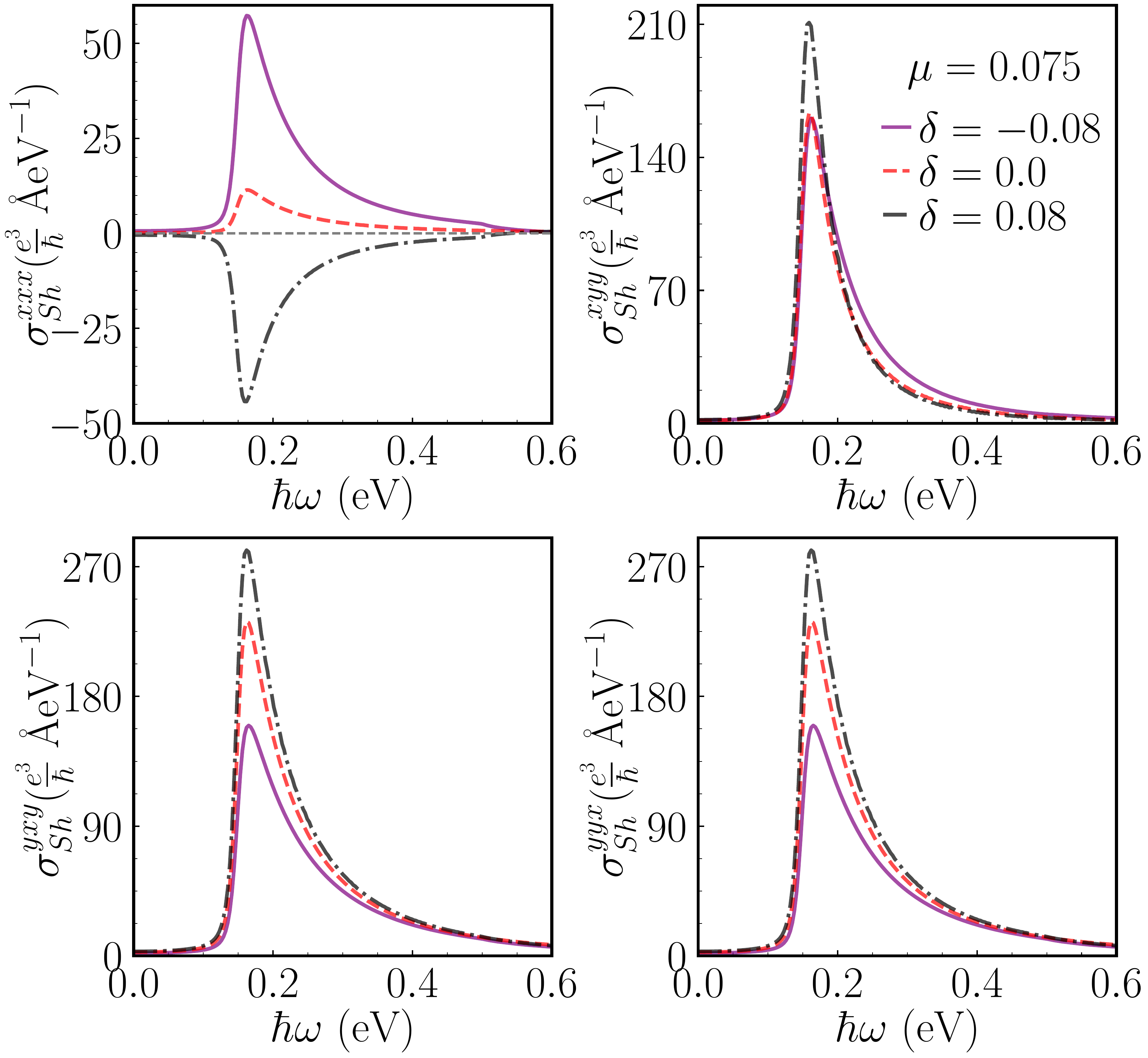}
    \caption{Perturbation-parameter ($\delta$) dependence of the LPGE shift current in the type-II semi-Dirac system. The shift-current components (a) $\sigma^{xxx}_{Sh}$, (b)$\sigma^{xyy}_{Sh}$, (c)$\sigma^{yxy}_{Sh}$, and (d)$\sigma^{yyx}_{Sh}$ are plotted as functions of photon frequency for different values of $\mu$. The $xxx$ component undergoes a sign reversal as $\delta$ is tuned, whereas the other three components remain strictly negative. Peak positions are largely insensitive to $\delta$, while amplitudes evolve smoothly, reflecting the geometric restructuring of the three-node type-II dispersion without substantial shifts in the dominant interband transition energy. The supporting model parameters are the same as in \fig{fig:type2_mu_var}.}
    \label{fig:type2_delta_var}
\end{figure}

In contrast to the type-I case, the type-II SD system hosts four nonvanishing shift-current components, $(xxx, xyy, yxy, yyx)$, each containing an odd number of $x$ indices, as dictated by the combined mirror symmetry $\mathcal{M}_y$ and time-reversal symmetry $\mathcal{T}$. The real parts of all surviving components display a single peak whose position shifts to higher frequencies while the peak magnitude decreases with increasing chemical potential $\mu$, mirroring the trend observed in the type-I system. This behavior is illustrated in \fig{fig:type2_mu_var}. Notably, the peak amplitude in the type-II case is consistently larger than that of the type-I SD system, reflecting the additional interband channels made available by its three-node topology. For $\delta = 0.08$, the $xxx$ shift-conductivity component displays a sign reversal compared to the remaining components $(xyy, yxy, yyx)$: specifically, $xxx$ is negative, whereas the other three components remain positive over the relevant frequency range.

Conversely, \fig{fig:type2_delta_var} shows the $\delta$-dependence of the real part of the shift conductivity in the type-II system. As $\delta$ is varied, the system undergoes a Lifshitz transition, and the $xxx$ component exhibits a clear sign reversal [\fig{fig:type2_delta_var}(a)], signaling a redistribution of $\tilde{\Gamma}^{abc}_{mp}$ associated with the change in the underlying node configuration. This sign change originates from the corresponding symplectic connection $\bar{\Gamma}^{xxx}_{+-}$, which changes sign upon reversing the sign $\delta$, as evident from \fig{fig:type2_symp_connection}(a–c). In contrast, the remaining three components $(xyy, yxy, yyx)$ retain a negative sign throughout all three phases, since their associated symplectic connections do not undergo a sign reversal upon tuning $\delta$. As an illustrative example, the behavior of $\bar{\Gamma}^{yyx}_{+-}$ is shown in Fig.~\ref{fig:type2_symp_connection}(d–f), where its sign structure remains unchanged across the Lifshitz transition. Together, the CPGE and LPGE responses provide a distinct optical fingerprint that clearly differentiates the type-I and type-II SD phases.

\section{Conclusion} \mylabel{sec:conclusion}
In summary, 
we have demonstrated that the photogalvanic effect provides a powerful quantum-geometric diagnostic of anisotropic semi-Dirac systems. Using quantum kinetic theory within the relaxation-time approximation, we systematically identified the injection, shift, resonance, higher-order pole, and anomalous contributions to the PGE conductivity and connected them to fundamental geometric quantities such as Berry curvature, quantum metric, metric connection, and symplectic connection. Our analysis reveals a clear contrast between type-I and type-II semi-Dirac phases. The CPGE, governed primarily by Berry curvature, is strongly enhanced in the type-II phase while remaining qualitatively similar in structure to the type-I case. In contrast, the LPGE shows distinct qualitative behavior controlled by the symplectic connection and symmetry constraints, which demonstrate a much larger overall response. Notably, the $xxx$ component in the type-II phase reverses sign across the Lifshitz transition upon tuning the perturbation parameter $\delta$, providing a direct optical marker of the topological change, while other components remain sign-invariant. These findings establish nonlinear photovoltaic and photocurrent responses as sensitive probes of SD band geometry and offer a unified framework for distinguishing between the two topological classes.

\noindent
The predicted signatures can be experimentally accessible in two-dimensional materials platforms—such as engineered oxide heterostructures, strained black phosphorus, and cold-atom optical lattices, where semi-Dirac phases have been proposed or realized. Polarization-resolved photocurrent measurements, photoconductivity spectroscopy, and pump–probe techniques operating in the terahertz to mid-infrared regimes can directly test our proposed frequency-dependent structures, sign reversals, and anisotropies of PGE conductivities. In addition, the large magnitude and good tunability of CPGE and LPGE responses in type-II SD systems make them potential candidates for technological applications in photodetection, directional current generation, nonlinear optical modulators, and valley-sensitive devices that exploit anisotropic light–matter coupling. Our results thus provide both a diagnostic tool for characterizing SD materials and a design principle for leveraging their nonlinear optical properties in next-generation optoelectronic technologies.
\section{Acknowledgement}
S.~N. acknowledges financial support from Anusandhan National Research Foundation (ANRF), Government of India via the Prime Minister's Early Career Research Grant: ANRF/ECRG/2024/005947/PMS. B.~G. is supported by INSPIRE, DST, Government of India.

\bibliography{refs}
\end{document}